\documentclass{elsart}
\usepackage{graphicx}
\usepackage{amsmath,amssymb}
  \def\mathcomposite{%
     \@ifstar
        {\def\@mathcomposite@option{%
            \baselineskip\z@skip\lineskiplimit-\maxdimen}%
         \@mathcomposite}%
        {\let\@mathcomposite@option\offinterlineskip
         \@mathcomposite}}
  \def\@mathcomposite{%
     \@ifnextchar[\@@mathcomposite{\@@mathcomposite[0]}}
  \def\@@mathcomposite[#1]#2#3#4{%
     #2{\mathchoice
        {\@mathcomposite@{#1}{#3}{#4}\displaystyle{1}}%
        {\@mathcomposite@{#1}{#3}{#4}\textstyle{1}}%
        {\@mathcomposite@{#1}{#3}{#4}%
         \scriptstyle\defaultscriptratio}%
        {\@mathcomposite@{#1}{#3}{#4}%
         \scriptscriptstyle\defaultscriptscriptratio}}}
\def\@mathcomposite@#1#2#3#4#5{%
     \vcenter{\m@th\@mathcomposite@option
        \dimen@\f@size\p@\dimen@#1\dimen@\dimen@#5\dimen@
        \divide\dimen@ 18
        \edef\@mathcomposite@skipamount{\the\dimen@}%
        \ialign{\hfil$#4##$\hfil\cr
           #2\crcr
           \noalign{\vskip\@mathcomposite@skipamount}%
           #3\crcr}}}

\def\sla#1{\rlap{\kern .13em /}#1}

\def\simge{
    \mathrel{\rlap{\raise 0.511ex
        \hbox{$>$}}{\lower 0.511ex \hbox{$\sim$}}}}
\def\simle{
    \mathrel{\rlap{\raise 0.511ex
        \hbox{$<$}}{\lower 0.511ex \hbox{$\sim$}}}}
\makeatother

\begin{document}
\begin{frontmatter}

\title{
Spectral Analysis of Excited Nucleons in Lattice QCD
with Maximum Entropy Method
}

\author[label1]{Kiyoshi Sasaki},
\ead{\\ \ \ \ \ ksasaki@ccs.tsukuba.ac.jp}  
\author[label2,label3]{Shoichi Sasaki},
\ead{\\ \ \ \ \ ssasaki@phys.s.u-tokyo.ac.jp}  
\author[label2]{Tetsuo Hatsuda},
\ead{\\ \ \ \ \ hatsuda@phys.s.u-tokyo.ac.jp}  
\address[label1]{Center for Computational Science, University of
Tsukuba, Tsukuba, Ibaraki 305-8577, Japan}
\address[label2]{Department of Physics, University of Tokyo, Tokyo 113-0033, Japan }
\address[label3]{RIKEN BNL Research Center, Brookhaven National Laboratory,
Upton, NY 11973-5000, USA }

\begin{abstract}

   We study the mass spectra of excited baryons
   with the use of the lattice QCD simulations.
   We focus our attention on
   the problem of the level ordering between the 
   positive-parity excited state $N'(1440)$ (the Roper resonance)
   and the negative-parity excited state $N^*(1535)$. 
   Nearly perfect
   parity projection is accomplished 
   by combining the quark propagators with periodic and anti-periodic boundary 
   conditions in the temporal direction.
   Then we extract the spectral functions from the lattice data by
   utilizing the maximum entropy method. 
   We observe that the masses of the $N'$ and $N^*$ states are close
   for wide range of the quark masses ($M_{\pi}=0.61-1.22$ GeV) 
   and for the physical point after the extrapolation in terms of the 
     quark mass.  The latter aspect is 
   in contrast to the phenomenological prediction of the phenomenological
   quark models. The role of the Wilson doublers in the baryonic
   spectral functions is also studied.

\end{abstract}

\end{frontmatter}

\section{Introduction}
\label{sec:Introduction}

   The lattice QCD simulations 
   have been shown to be  a  
   powerful tool to study the properties of hadrons from first 
   principle. In particular, ground state hadron masses
   in the 
   quenched approximation agree with the 
   experimental values in 10\% accuracy \cite{Aoki:2002fd}.
   Furthermore unquenching of the dynamical quarks tends to improve
   the agreement \cite{Ishikawa:2004xq}. On the other hand,
   studies of the  excited hadrons are still
   in an exploratory  stage in spite of their physical importance. 
   (For a brief review, see \cite{Sasaki:2003xc} and references therein.)

  In this paper, from the quenched lattice QCD simulations
  with Wilson fermions,
  we study excited nucleons with a special focus on the spectrum
  in the positive-parity channel
  $N'(1440)$ (the Roper resonance) and that in
  the negative-parity channel $N^*(1535)$.
   
  The fact that the mass of the $N'(1440)$ is smaller than
  that of the $N^*(1535)$ has been a long standing puzzle:
  Indeed,  phenomenological quark models predict that
  $M_{N'}-M_N \sim 2 \times (M_{N^*}-M_N)$  
  because $N=(1s)^3$ while $N'=(1s)^2(1p)$ and 
  $N^* = (1s)^2(2s)$ or $(1s)(1p)^2$~\cite{Capstick:2000qj}.
  As possible resolutions of this puzzle,   
  exotic descriptions of the $N'$ state and/or the $N^*$ state
  have been proposed such as 
  the chiral quartet scheme~\cite{Jido:1999hd} 
  and the pentaquark picture~\cite{Jaffe:2003sg} 
  as well as the quark model with
  the effective flavor-spin interaction~\cite{Glozman:1995fu}.

  In principle,  the lattice QCD simulation ought to give the 
  final answer to this problem.  However,
  definite conclusions   
  have not been drawn yet even in the quenched lattice QCD studies
  because of several technical reasons: First of all, one needs to
  make appropriate projection and isolation of the excited states
  to differentiate the $N$, $N'$ and $N^*$ states.  
  Furthermore, lattice artifact due to finite lattice volume
  becomes severer for the excited states than the ground 
  state ~\cite{{Sasaki_Roper},{Sasaki:2005ug}}.
  
  To overcome these problems, 
  (i) we adopt a technique to combine the quark propagators
  with periodic and anti-periodic boundary conditions in time   
  for precise parity projection \cite{Sasaki:2001nf},
  (ii) we employ
  the maximum entropy method (MEM) \cite{MEM_lQCD,Sasaki_Roper}
  for extracting the excited mass spectra, and
  (iii) we utilize
  a  larger spatial size  ($La \simeq 3.0$ fm) with
  the lattice spacing $a \simeq 0.093$ fm ($\beta=6.0$).
 
 Our main conclusion is that $N'$ and $N^*$ states 
 are nearly degenerate within statistical errors in the wide range 
 of the quark mass ($M_\pi=0.61-1.22$ GeV).
 We found that, after the chiral extrapolation with an empirical ``curve fit" formula, 
 the masses of the $N'$ and $N^*$ states are consistent with 
 the approximate degeneracy observed in experiment.   
 Also, we found a firm evidence of high-lying bound states
 composed of  Wilson doublers.
  
 This paper is organized as follows.
 In Section~\ref{sec:Numerical_methods},
 we provide a brief introduction to 
 the parity projection and MEM.
 After summarizing our simulation
 parameters in Section \ref{sec:Simulation_parameters},
 we discuss the numerical results obtained from the MEM analysis
 of the spectral function in Section \ref{sec:NSTR_and_Roper}.
 The bound states of Wilson doublers which appear
 as high energy peaks in the spectral function are discussed
 in Section~\ref{sec:Bound_state_of_doublers}. 
 A brief comparison with previous studies is given
 in Section \ref{sec:Comparison_with_other_results}. 
 Section~\ref{sec:Conclusion} is devoted to 
 summary and concluding remarks.

\section{Parity projection and MEM}
\label{sec:Numerical_methods}

The nucleon is described by the trilinear quark composite operator given by
%
%
\begin{eqnarray}
N(x)=\varepsilon_{abc}[u^{\rm T}_a(x)C\gamma_5 d_b(x)]u_c(x),
\label{eq:conv-N}
\end{eqnarray}
which is composed of a local scalar-diquark operator and a spectator-like quark field.
This local composite operator couples to  both positive and negative parity states
~\cite{{Fucito:1982ip},{Sasaki:2001nf}}.
The asymptotic form of the zero-momentum  nucleon correlation,  $G(t)$,
may be expressed in finite temporal extent, $T$, as
%
%
\begin{eqnarray}
  G(t) &=& \sum_{\vec{x}}\langle  N(\vec{x},t)\bar{N}({\vec 0},0) \rangle 
  \nonumber \\
  & \longrightarrow  & \left[\Lambda_{+} A_{+}\mathrm{e}^{-M_+t}
           -\Lambda_{-} A_{-} \mathrm{e}^{-M_-t}\right]
 \nonumber \\
 & & \ \ \ \ \
  + b\left[\Lambda_{-}  A_{+} \mathrm{e}^{-M_+(T-t)}
           -\Lambda_{+} A_{-} \mathrm{e}^{-M_-(T-t)}\right]
\label{eq:gt}
\end{eqnarray}
 with $\Lambda_{\pm} = (1 \pm \gamma_4)/2$. The coefficient 
 $b=1 (-1)$ is for the periodic (anti-periodic) boundary condition
 in the temporal direction.
 $M_+ (M_-)$ is the ground state mass of the positive (negative) parity
 nucleon. The second and forth terms on the right hand side of Eq.(\ref{eq:gt})
 appear since the nucleon is a composite particle.
 Due to the presence of the third and forth terms corresponding to 
 the primal reflections from the time boundary,
 the naive projection $\mathrm{Tr}\left( \Lambda_{\pm} G(t)\right)$
 is insufficient to extract a contribution of the desired 
 parity state.~\footnote{ 
 The most primitive prescription to avoid the contamination from
 the opposite parity states is to analyze the data only up to about
 $T/2$.}
 A solution to this problem is to realize the free-boundary 
 situation ($b=0$), which can be achieved
 by using a linear combination of the quark propagators 
 with periodic and anti-periodic boundary conditions 
 in time~\cite{Sasaki:2005ug,Sasaki:2001nf}. 
 The linear combination in the quark level automatically 
 fulfills the linear combination in the hadronic level 
 and then manages the free-boundary situation
 up to the first wrap-round effect~\cite{Sasaki:2005ug}.
 Adopting this procedure, 
 one can project out the desired  
 parity state by $\Lambda_{\pm}$ accurately without
 the contamination from the opposite parity state.
 
The baryonic correlation $G(t)$ with the above parity projection 
has a spectral representation,
%
%
\begin{equation}
  G(t)\equiv \int_0^\infty d\omega\  K(t,\omega)A(\omega), 
  \label{eqn:GKA}
\end{equation}
where $K(t,\omega)=\exp(-\omega t)$ is the Laplace integral kernel.
 The spectral function (SPF), $A(\omega)$,  carries all the information of the 
 hadronic states (both ground and excited states).  
 The discretized version of Eq.~(\ref{eqn:GKA}) reads
 $G(t_i) = \sum_l \exp(-\omega_l t_i) A(\omega_l)$:
 We need values of $A(\omega_l)$ for sufficient number of 
 frequencies $\omega_l$, typically of $\mathcal{O}(100)$. 
 The Monte Carlo data $G(t_i)$ is, however, available
 only at some dozens of time slices on the lattice.
 This is a typical ill-posed problem, 
 where the number of equations is much smaller 
 than the number of unknown quantities.
 The maximum entropy method
 (MEM) introduced
 in Ref. \cite{MEM_lQCD}  enables us to determine SPF uniquely
 by supplementing the lattice data with prior knowledge 
 such as the positivity and
 the asymptotic behavior of SPF. 

The conditional probability of  $A(\omega)$ 
given the Monte Carlo data $G(t)$ and 
the prior knowledge $H$ is given by $P[A|GH]$ which is
evaluated by using the Bayes' theorem as 
%
%
\begin{equation}
  P[A|GH] \propto P[G|AH]\ P[A|H] .
  \label{eqn:Bayes_Theorem}
\end{equation}
 Here $P[G|AH]$ is the standard likelihood function
 and $P[A|H]$ is given by the Shannon-Jaynes entropy:
%
%
\begin{equation}
  P[A|H] \propto \exp\left[\alpha 
  \sum_{l=1}^{N_\omega}
  \left[ A_l-\bar{A}_{l}-A_l \ln(A_l/\bar{A}_{l}) 
  \right]\right].
  \label{eq:PAH}
\end{equation}
 Given $G(t)$, $\alpha$ and $\bar{A}$, the most probably $A$  is obtained 
 by  the stationary condition, $(\delta/\delta A) P[A|GH] =0 $, which 
 has a unique 
 solution if it exists \cite{MEM_lQCD}.  
 Also, the reliability of the solution $A$ can be tested by
 the second variation, $(\delta/\delta A)^2 P[A|GH] $.
 We have $\bar{A} \equiv m_0 \omega^n $ ($n= 2 (5)$ for meson (baryon))
 in Eq.(\ref{eq:PAH}): This is called  the {\it default model}
 in which the parameter $m_0$ may be fixed either by the 
 perturbative asymptotic behavior of SPF at large $\omega$, 
 or by demanding that the resultant $A$ has smallest 
 error bar.
 By evaluating the probability  $P[\alpha|GH]$ using the 
 Bayes' theorem, 
 final form of the spectral function is obtained
 by averaging over $\alpha$ with the weight factor $P[\alpha|GH]$. 
 For more details on MEM techniques, see Ref.~\cite{MEM_lQCD}.

\section{Simulation parameters}
\label{sec:Simulation_parameters}

%
%
\begin{table}[bth]
\bigskip
\caption{Simulation parameters in this study.
  $\beta=6/g^2$, $L$ ($T$) is the number of the spatial (temporal) site,
  $a$ ($La$) is the lattice spacing (spatial lattice size) in the 
  physical unit, and $N_\mathrm{conf}$ is the number of
  gauge configurations.  The Sommer parameter $r_0=0.5$ fm is used to
  fix the scale~\cite{Guagnelli:1998ud}. }
\begin{center}
\begin{tabular}{lccclc}
\hline
$\beta$ &
$L^3 \times T$ & 
$a$ (fm)&
$La$ (fm) &
hopping parameter &
$N_\mathrm{conf}$ \\
\hline
5.8 & $24^3\times 32$ & 0.136 &  3.3
& 0.1600, 0.1570, 0.1555, 0.1540, 0.1530 & 200\\
6.0 & $16^3\times 32$ & 0.093 & 1.5 
& 0.1550, 0.1530, 0.1515 & 580 \\
    & $32^3\times 32$ & 0.093 & 3.0 
& 0.1550, 0.1530, 0.1515, 0.1500  & 200\\
\hline
\end{tabular}
\end{center}
\label{table:Simulation_parameters}
\bigskip
\end{table}

We have performed simulations in quenched QCD
with the single plaquette gauge action 
and the Wilson fermion action. 
We utilized the Metropolis algorithm
with 20 hits at each link update. 
The gauge ensembles in each simulation are separated 
by 1000 sweeps for $\beta=6.0$, and 600 sweeps for $\beta=5.8$.
For the matrix inversion, we used BiCGStab algorithm~\cite{Frommer:1994vn} 
with the convergence condition $|r|< 10^{-8}$ for residues.
We calculate the point-to-point nucleon propagator constructed
by the point-to-point quark propagator so that the 
spectral function is assured to be positive semi-definite.
Our calculation has been performed on a Hitachi SR8000 parallel computer 
at KEK (High Energy Accelerator Research Organization), 
using the extended code based on the Lattice Tool Kit (LTK)\cite{Choe:2002pu}.

 Our simulation parameters 
 are listed in Table~\ref{table:Simulation_parameters}. 
 Physical scale is set by 
 the Sommer parameter $r_0=0.5$ fm~\cite{Guagnelli:1998ud}. 
 Our main conclusion will be drawn from the data taken
 on the finest and largest lattice ($\beta=6.0$ and $L=32$) 
 in the Table~\ref{table:MEM_ALL_RESULTS}.
 The case with coarser lattice spacing ($\beta=5.8$ and $L=24$)
 is also studied to analyze the nature of the high-lying
 peaks of the spectral function associated with
 the Wilson doublers.
 A supplementary data with a smaller lattice size
 ($L=16$ with $\beta=6.0$) are also taken
 to check the effect of finite lattice size on the 
 excited nucleon spectrum. 
 For each data, several hopping parameters are used 
 to cover the range $ M_\pi \simeq 0.6-1.2$ GeV. 
 
\section{$N^*(1535)$ and the Roper resonance $N'(1440)$}
\label{sec:NSTR_and_Roper}

%
%
\begin{figure}[tbhp]
\bigskip
  \begin{center}
  \includegraphics[height=80mm]{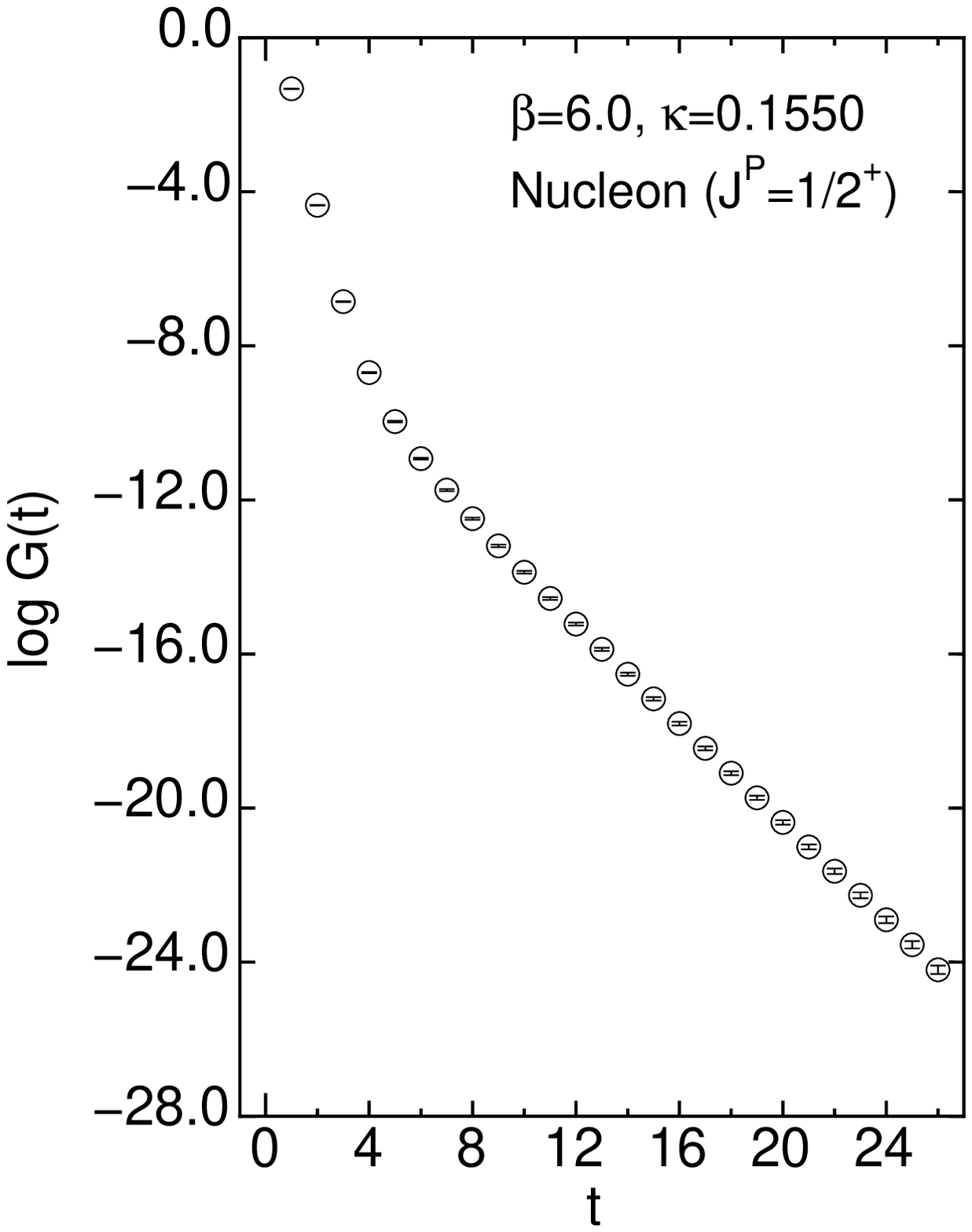}
  \hspace{10mm}
  \includegraphics[height=80mm]{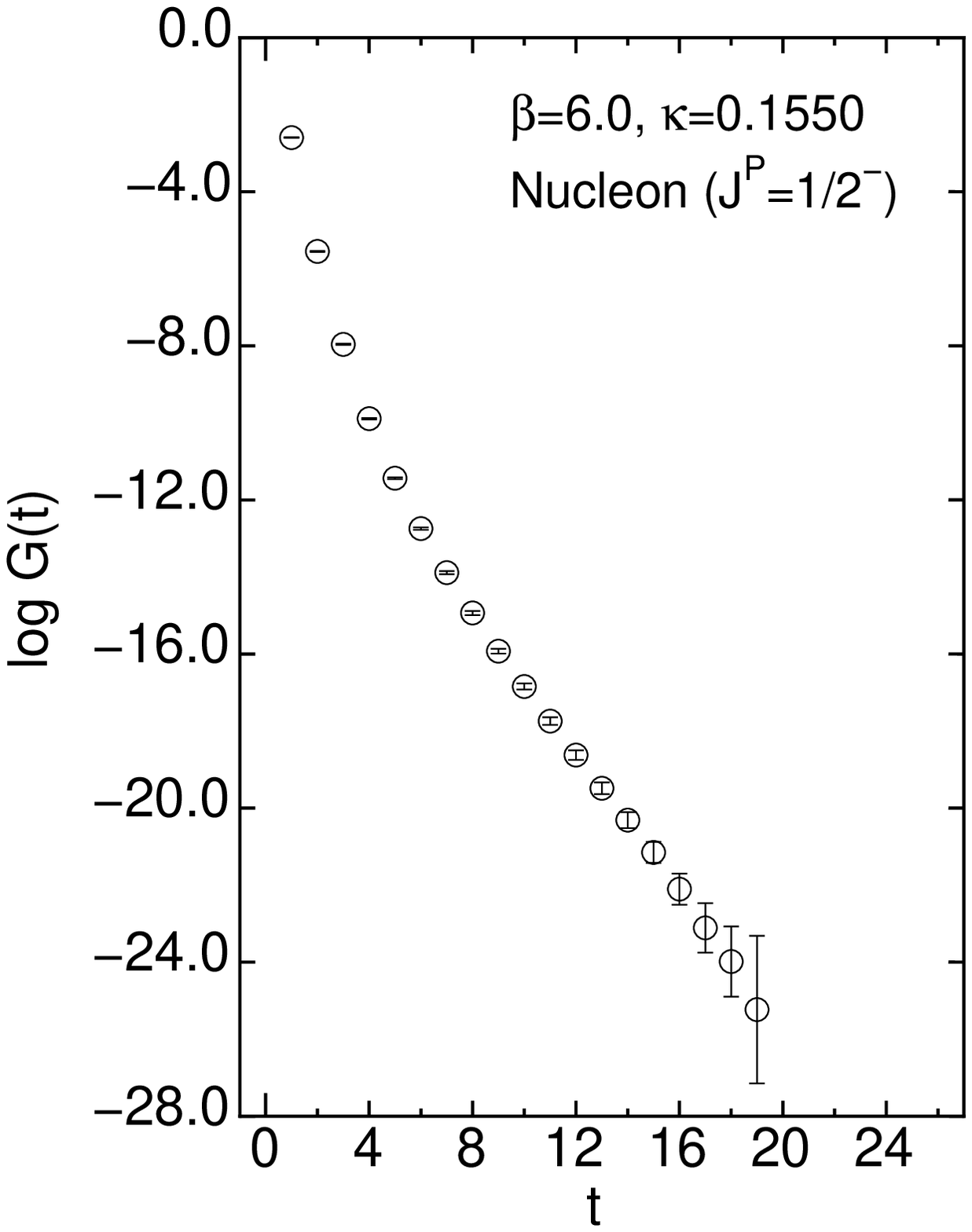}
  \end{center}
  \caption{Examples of the correlation function $G(t)$
   generated by the Monte Carlo simulations
   as a function of Euclidean time $t$ 
  for $\beta=6.0$, $\kappa=0.1550$ and  $L=32$. 
  The left panel corresponds to the positive-parity nucleon channel 
  and the right panel to the negative-parity nucleon channel. 
In the right panel, we have not shown the data for $t\ge 20$, 
since the statistic errors become much larger 
than the mean values 
and the data do not have statistical significance.}
  \label{fig:NUCL1_b6.0_32x32x32x32_CORRELATION}
\bigskip
\end{figure}
 %

 %
%
\begin{figure}[tbhp]
\bigskip
  \begin{center}
  \includegraphics[height=50mm]{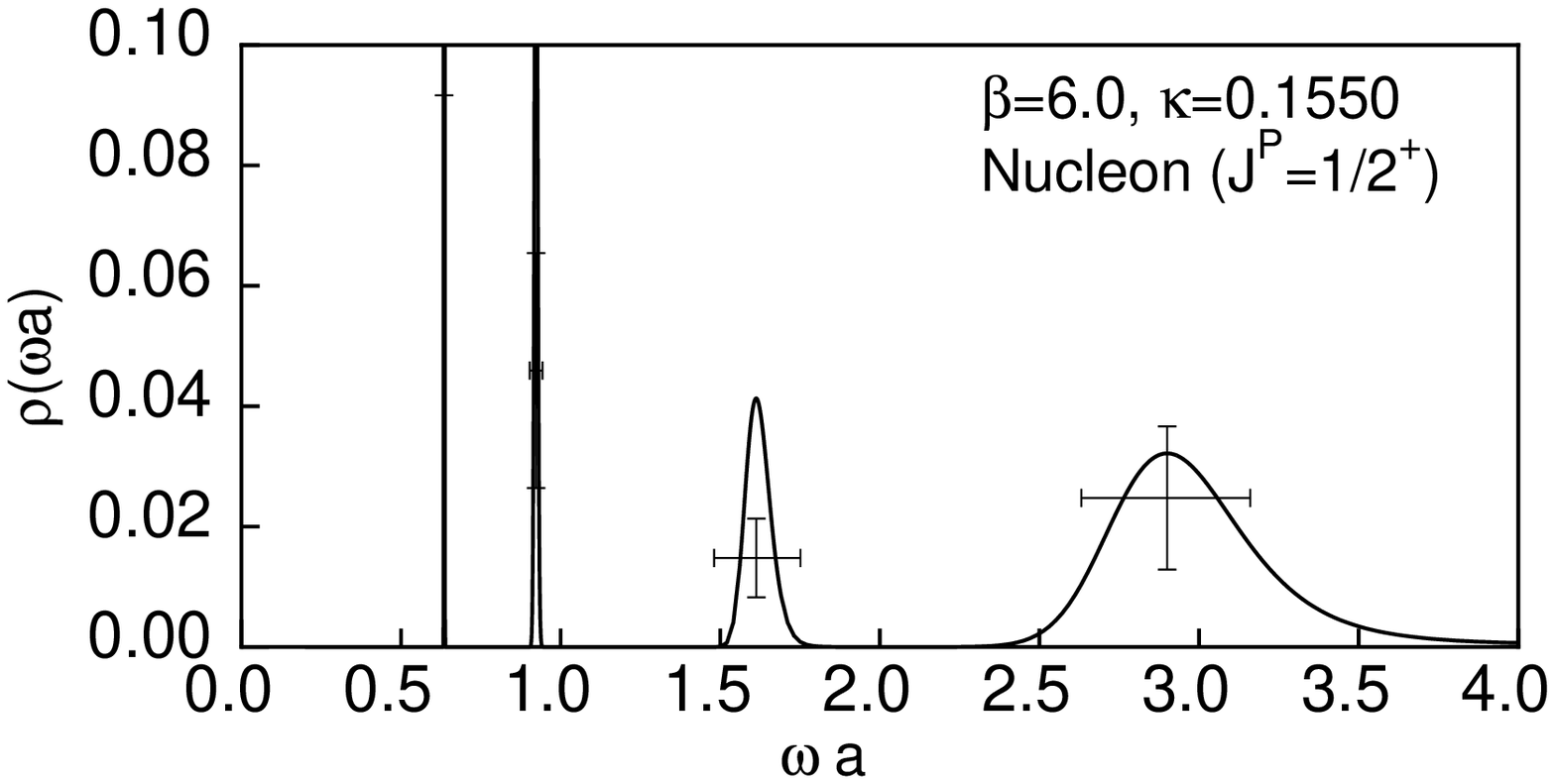}
  \includegraphics[height=50mm]{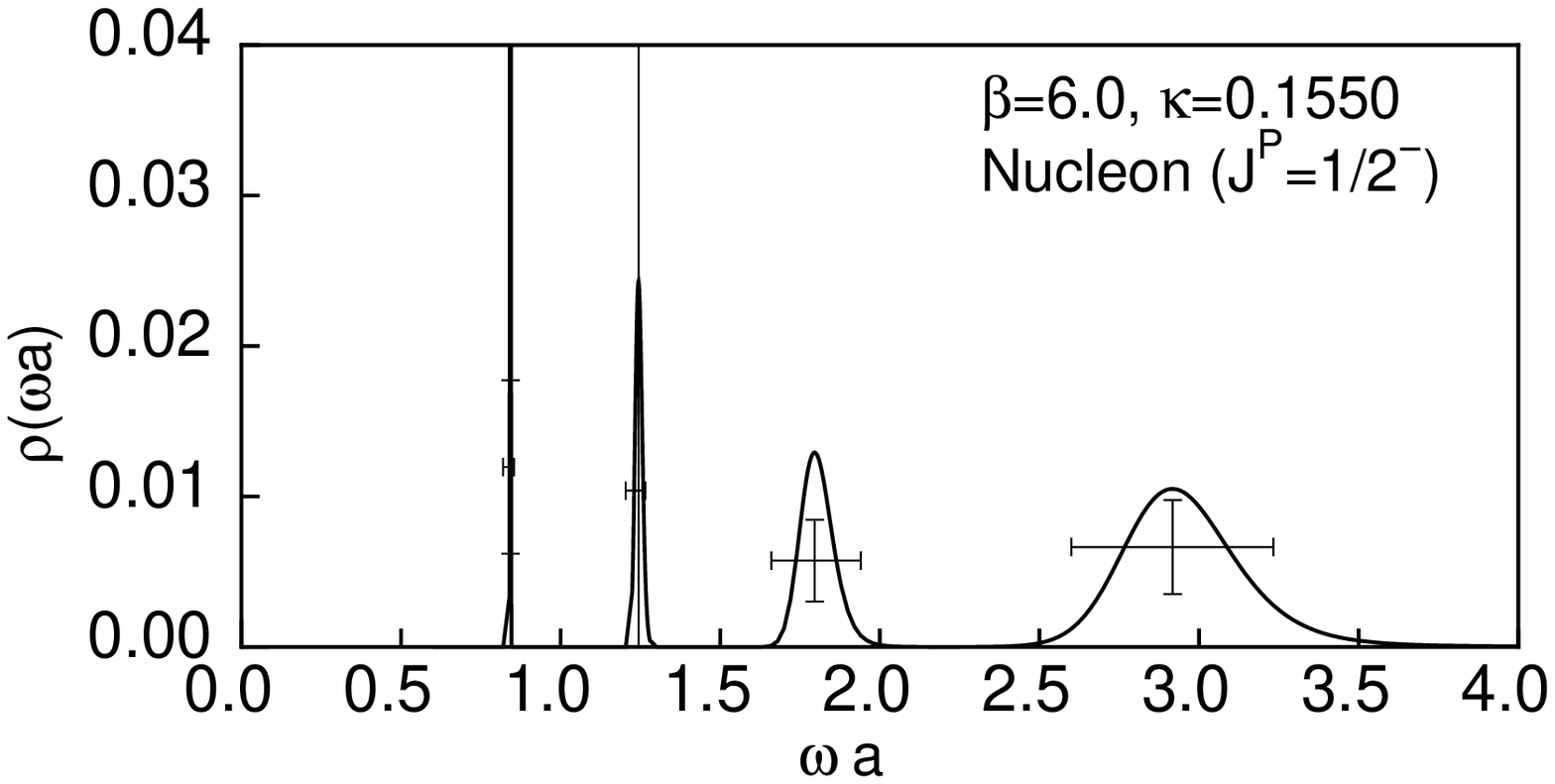}
  \end{center}
  \caption{The dimensionless spectral function, $\rho(\omega)=A(\omega)/\omega^5$, 
  as a function of the dimensionless frequency $\omega a$
  for $\beta=6.0$, $\kappa=0.1550$ and  $L=32$. 
  The upper panel corresponds to the positive-parity nucleon channel 
  and the lower panel to the negative-parity nucleon channel. }
  \label{fig:NUCL1_b6.0_32x32x32x32_SPECTRAL}
\bigskip
\end{figure}

  In Fig.\ref{fig:NUCL1_b6.0_32x32x32x32_CORRELATION},
   we have shown examples of the correlation functions $G(t)$ 
  in the positive parity and negative channels 
 for $\beta=6.0$ ($a=0.093$ fm),
  $\kappa=0.1550$ ($M_{\pi}\simeq 0.6$ GeV) and $L=32$ ($La\simeq 3.0$ fm).
 Fig.\ref{fig:NUCL1_b6.0_32x32x32x32_SPECTRAL}
 show the dimensionless spectral functions, 
  $\rho(\omega)\equiv A(\omega)/ \omega^5$, 
 obtained from the MEM analysis of $G(t)$ 
  in Fig.\ref{fig:NUCL1_b6.0_32x32x32x32_CORRELATION}.
  The default model in the MEM analysis for the nucleon channel
  is chosen to be about $m_0=0.004$, which is deduced from 
  the asymptotic form of the SPF at the one-loop level in the 
  perturbative QCD in the continuum~\cite{Sasaki_Roper}.
  We have check that the final results are
  insensitive to the choice of $m_0$ around this value.
  The temporal interval of the Monte Carlo data $G(t)$, 
  used in the MEM analysis, is chosen to be
  $[ t_\mathrm{min}, t_\mathrm{max}]=[1, 20]$. 
  Exceptions are for the negative-parity nucleon with light quark masses
  ($\kappa=0.1530,~ 0.1550$) in which $t_\mathrm{max}=19$
 is used because of the noisy signals for $t\ge 20$.  
  Note that $t=0$ corresponds
  to a source location and is excluded in the analysis. 
  It is worth mentioning that sufficiently large statistics makes
  the reconstructed image of SPF stable against variation of $t_\mathrm{max}$
  beyond some critical point around $t_\mathrm{max}\approx18-19$. 
  In our MEM analysis, we have chosen the region, $0.2/a\le\omega \le 2\pi/a$,
  to cover all the relevant states available on our lattice.
  We have carefully
  checked that changing the lower (upper) limit of $\omega$ to smaller (larger)
  values have no effect on the finial results.  
 
  The crossbars attached to the SPFs in 
  Fig.\ref{fig:NUCL1_b6.0_32x32x32x32_SPECTRAL} are attached 
  with the procedure given in \cite{MEM_lQCD}.
  First of all, we arbitrarily choose an interval in the 
  frequency space so that it covers one of the peaks 
  in SPF.  The horizontal bar in the figure
  represent this interval and is not at all related to 
  the uncertainly of the peak position.
  After setting the interval, the uncertainty of the peak
  height averaged over the interval can be calculated
  from the second variation of $P[A|GH]$ with respect
  to $A$.  The vertical bar in the figure 
  represents $\pm 1 \sigma$ error of such averaged peak height. 
  Fig.\ref{fig:NUCL1_b6.0_32x32x32x32_SPECTRAL} shows that 
  the peaks are statistically significant
  except for the marginal second peak in the lower panel.
  We identify the first (second) peak of the upper panel 
  in the figure as $N$ ($N'$), and the first peak in the lower
  panel as $N^*$.

  Carrying out the MEM analysis in the meson and baryon channels
  for different values of $\kappa$ and identifying the 
  positions of low-lying peaks, we 
  extracted the hadron masses as  
  summarized in Table \ref{table:MEM_ALL_RESULTS}. 
  The statistical errors in the parentheses are evaluated by the jack-knife method. 
  Critical hopping parameter $\kappa_c$
  at which the pion mass vanishes reads $\kappa_c=0.15668(21)$
  for $\beta=6.0$.
 Our simulations are done for quark masses which are
  not very close to the chiral limit so that
 a guidance by the chiral perturbation theory is 
 no longer useful for the extrapolation of hadron masses 
 down to the physical point.
 Furthermore, the leading chiral behavior of 
 the excited baryon masses is not known at present.
  In this situation, adopting a simple formula, 
 which can describe the observed dependence of the quark mass well, 
 is a reasonable choice to make the chiral extrapolation. 
 In this paper, we utilize the ``curve fit"; 
  %
  %
  \begin{equation}
  (a M_H)^2 = d_0 + d_2 (a M_\pi)^2. 
  \label{eqn:curve_fit}
  \end{equation}
  which does not include a term linear in $aM_{\pi}$, which is 
  responsible for the leading chiral behavior for the nucleon 
  in quenched chiral perturbation theory~\cite{Labrenz:1996jy}.
  As reported in Refs. \cite{Gockeler:2001db,Sasaki:2005ug}, 
  Eq.(\ref{eqn:curve_fit}) gives a better fit to the data 
  in the heavy-quark region than the linear fit.
 Note that we have not made exhaustive study on the chiral extrapolation 
 in this paper, since the systematic errors stemming from the finite-size effect 
 is more significant than those from the chiral extrapolation.
%
%
\begin{table}[b]
\bigskip
\caption{Masses of the pion, the $\rho$-meson,  the nucleon $N$ ($J^P=1/2^+$), 
the Roper $N'$ ($J^P=1/2^+$) and the $N^*$ ($J^P=1/2^-$) 
extracted from the data on the lattice with $L=32$ and $\beta=6.0$
by the MEM analysis.
The statistical errors are evaluated by the jack-knife method.
The values in the chiral limit are evaluated by the curve fit.}
\begin{center}
\begin{tabular}{c|ccccc}
\hline
$\kappa$   & $a M_\pi$  & $a M_\rho$  & 
$a M_N$    & $a M_{N'}$ & $a M_{N^*}$ \\
\hline
$\kappa_c$ & 0          & 0.318(15)   & 
0.464(21)  & 0.76(16)   & 0.689(102)    \\
0.1550     & 0.288(14)  & 0.422(13)   & 
0.635(14)  & 0.92(13)   & 0.843(68)   \\
0.1530     & 0.420(7)   & 0.505(4)    & 
0.791(9)   & 1.09(8)    & 1.030(56)   \\
0.1515     & 0.502(3)   & 0.568(3)    & 
0.898(7)   & 1.20(6)    & 1.143(50)   \\
0.1500     & 0.578(2)   & 0.630(5)    & 
1.000(6)   & 1.31(6)    & 1.236(49)   \\
\hline
\end{tabular}
\end{center}
\label{table:MEM_ALL_RESULTS}
\bigskip
\end{table}
%

\begin{figure}[tbhp]
\bigskip
\begin{center}
  \includegraphics[height=80mm]{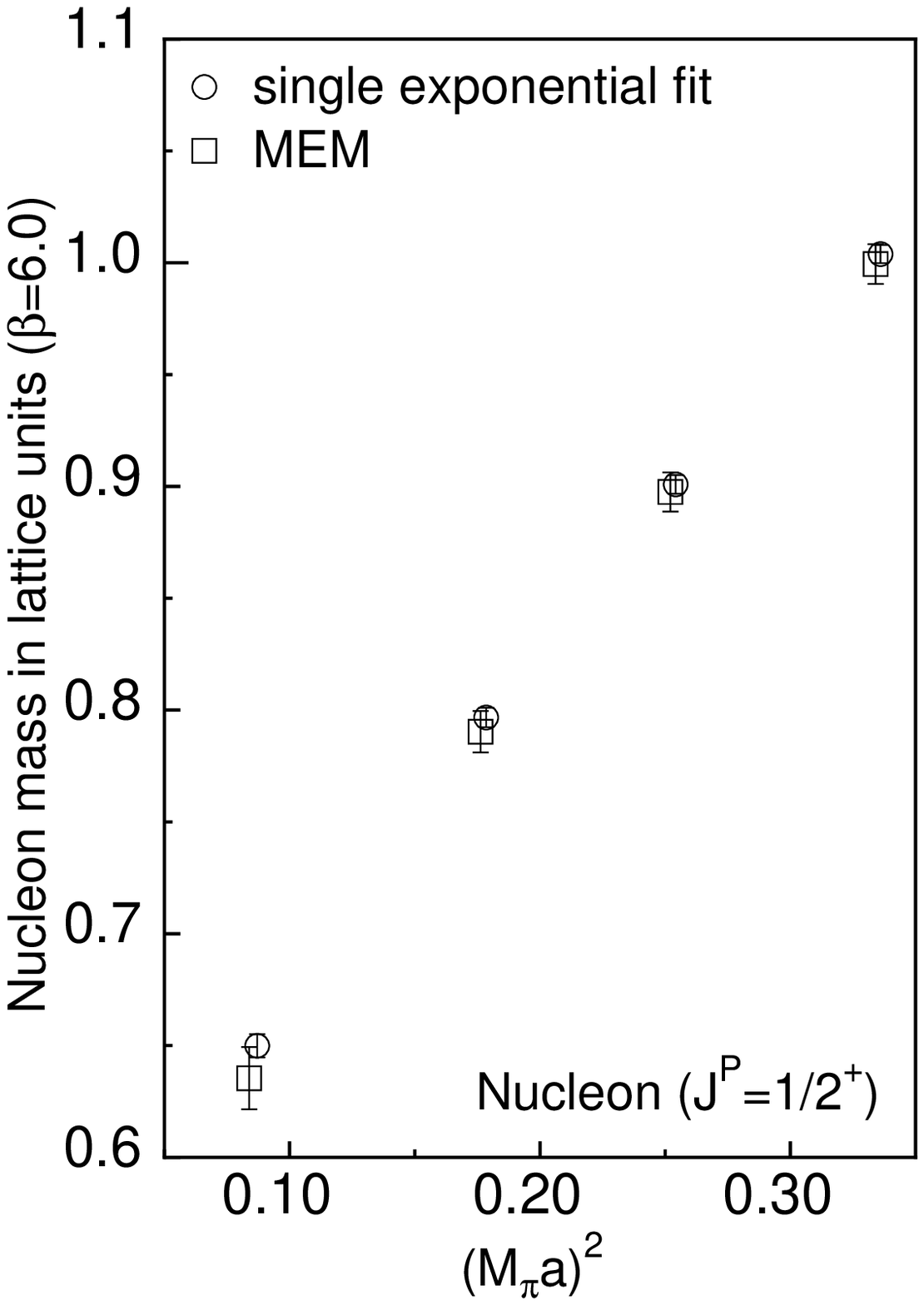}
  \hspace{20mm}
  \includegraphics[height=80mm]{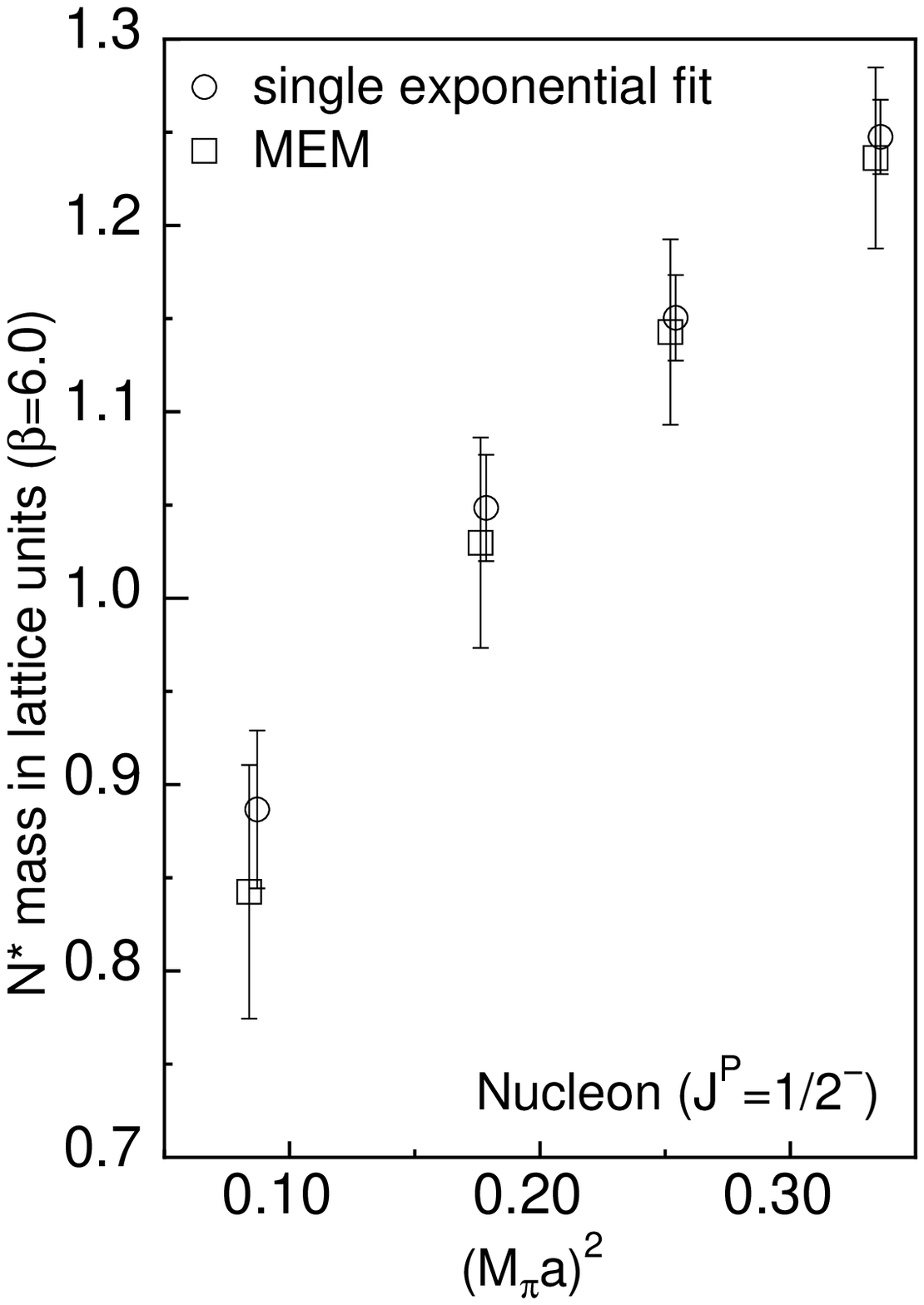}
\end{center}
\caption{A comparison between the position of the first peak 
  in the SPF obtained by the MEM analysis 
  and the  mass evaluated 
  by the single exponential fit, for $L=32$ and $\beta=6.0$.
  The left panel correspond to the positive-parity nucleon channel 
  and the right panel to the negative parity nucleon channel.}
\label{fig:P2_NUCL_NSTR_b6.0_32x32x32x32}
\bigskip
\end{figure}

  As for the ground state in each quantum number, one may compare
  the MEM result and the 
  conventional single exponential fit using $G(t\rightarrow {\rm large})$.
  Shown in Fig.\ref{fig:P2_NUCL_NSTR_b6.0_32x32x32x32}
  are the comparisons of the masses of the $N$ (the ground state in the positive-parity nucleon)
  and the $N^*$ (ground state in the negative-parity nucleon)
  in two approaches:
  The circles represent 
  the  mass evaluated by the single exponential fit, 
  while the squares represent 
  the position of the first peak  obtained by the MEM analysis.
  They are consistent with each other within the statistical error for
  the whole range of the quark masses.

%
%
%


%
%
\begin{figure}[tbhp]
\bigskip
  \begin{center}
  \includegraphics[height=50mm]{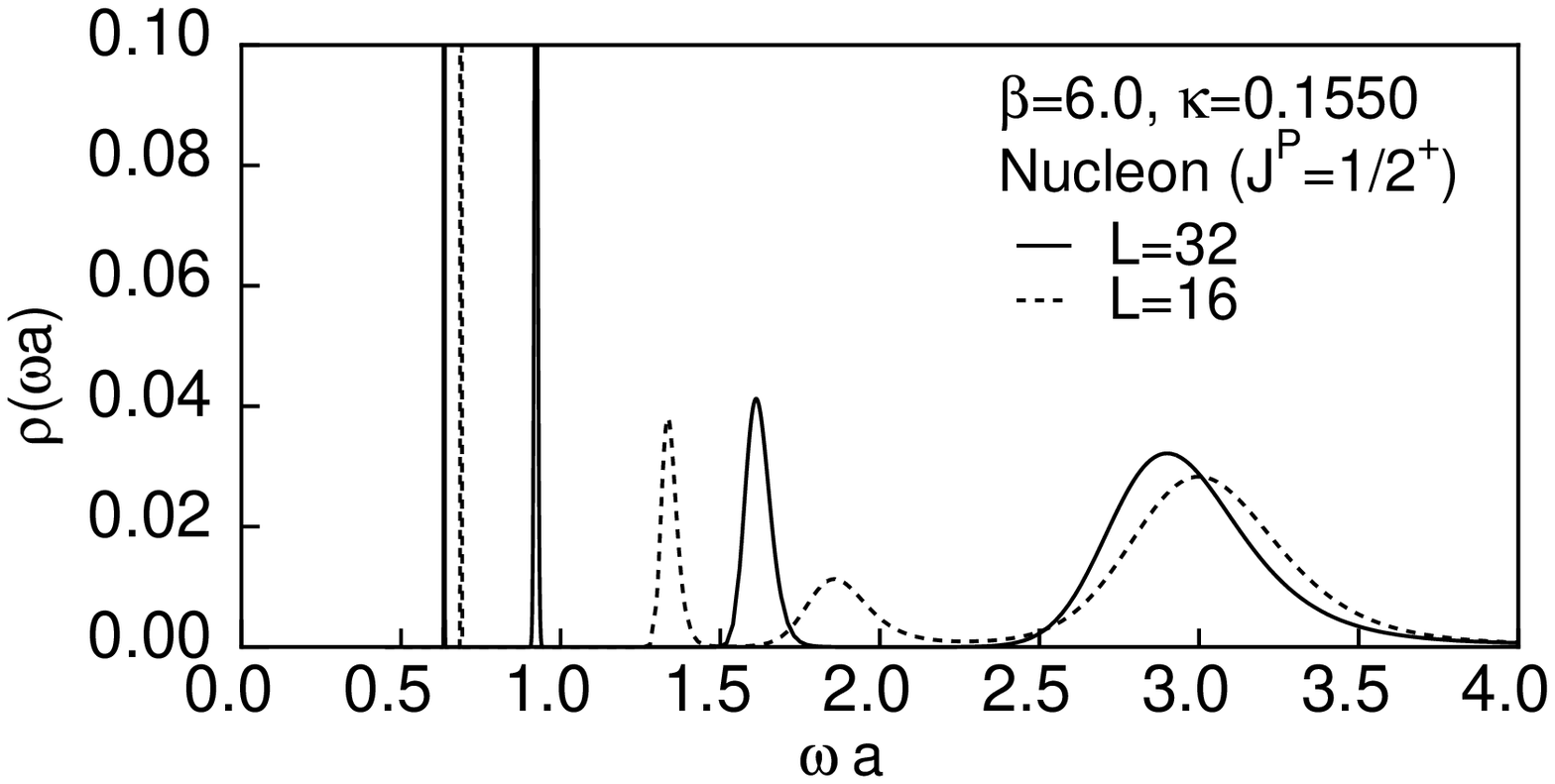}
  \includegraphics[height=50mm]{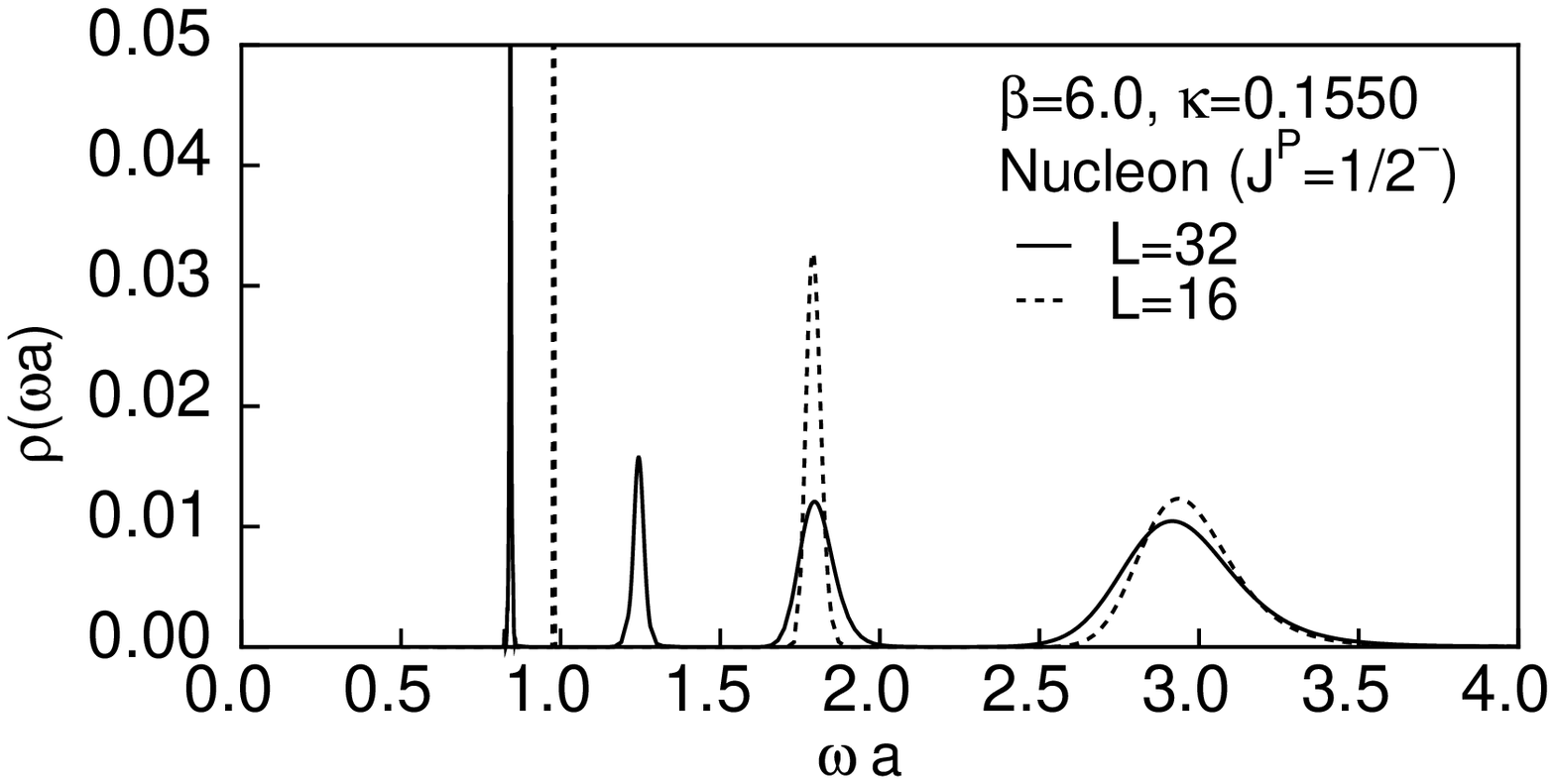}
  \end{center}
  \caption{The finite size effect on the SPF for  
  $\beta=6.0$ and $\kappa=0.1550$.   The upper (lower)
  panel corresponds to the positive (negative) parity nucleon channel.
  In each panel, 
  we compare the SPF obtained from the larger size, $L=32 (La \simeq 3.0$ fm),
  with that from the smaller size, $L=16 (La\simeq 1.5$ fm). 
  Note that the 2nd lowest peak on $L=32$ lattice
  disappears on $L=16$ lattice in the negative parity channel. 
 }
  \label{fig:NUCL_NSTR_b6.0_32x32x32x32_FSE}
\bigskip
\end{figure}

 Let us  consider  the finite size effect
 on the peaks of the spectral functions.
 In Fig.~\ref{fig:NUCL_NSTR_b6.0_32x32x32x32_FSE}, 
 we compare the SPFs on a larger lattice ($La\simeq 3.0$ fm)
 shown by the solid lines  
 with those on a smaller lattice ($La\simeq 1.5$ fm) shown
 by the dashed lines. 
 Although the position of the $N$ (the first peak in the upper
 panel) has a minor shift, the positions of the $N'$ (the second peak
 in the upper panel) and the $N^*$ (the first peak in the lower
 panel) have significant upward shifts as we decrease the lattice volume.
 This is consistent with an intuitive observation that
 the size of excited hadron is larger than that of the 
 ground state and thus receives a large finite volume
 effect.  Detailed analysis on the finite size effect on the $N$ and the $N^*$ 
 by two of the present authors shows that
 the spatial size of about 3 fm is necessary to obtain reliable
 results  even for relatively heavy quark masses 
 ($M_{\pi}\sim 1$ GeV)~\cite{Sasaki:2005ug}.
 With these observations, we will use the data for $La\simeq 3.0$ fm
 in this paper for final results.


%
%
\begin{figure}[tbhp]
\bigskip
  \begin{center}
  \includegraphics[height=100mm]{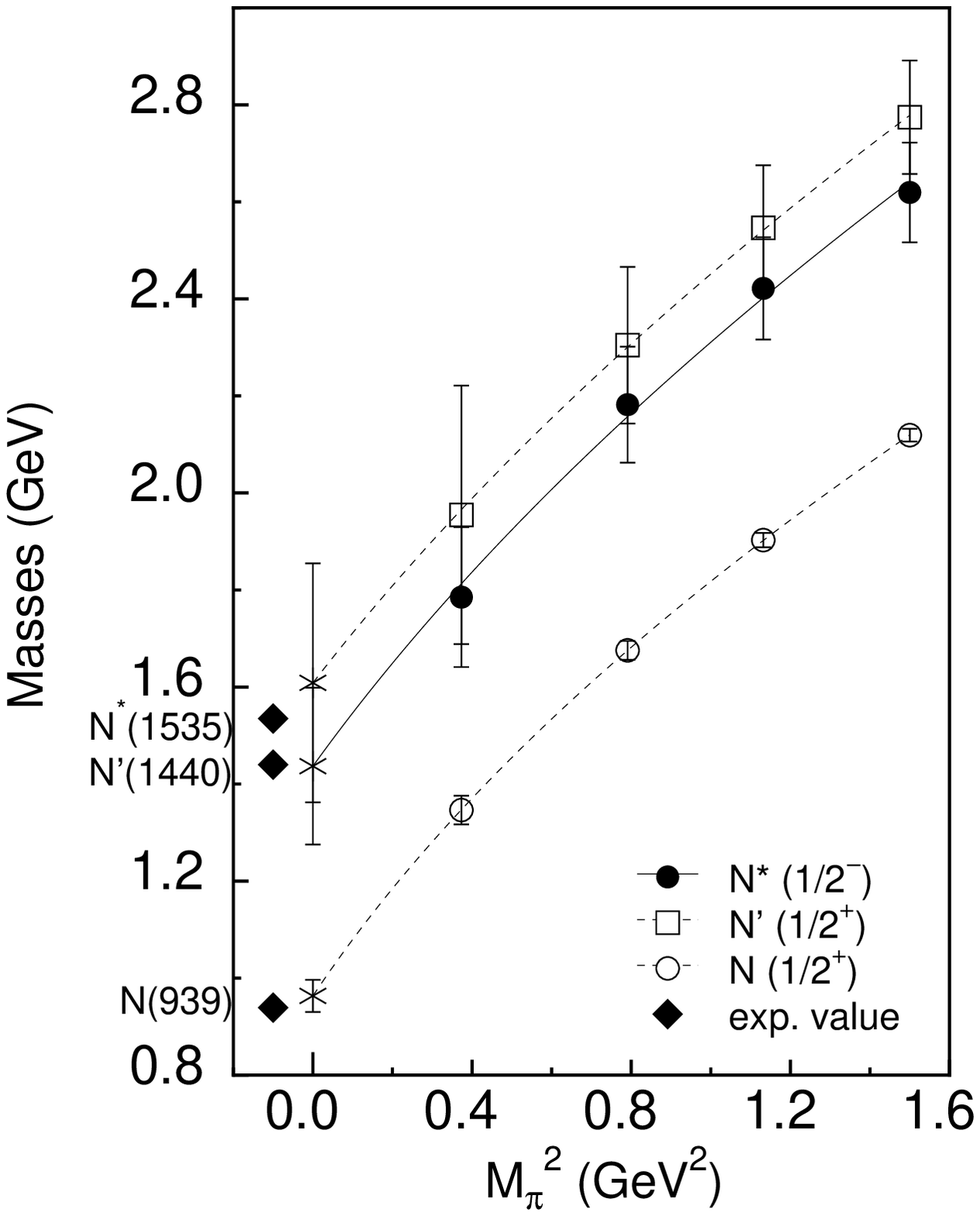}
  \end{center}
  \caption{
    Ground and excited nucleon spectra as a function of
    the pion mass squared in the physical unit. The experimental
    values for $N(940)$, $N'(1440)$ and $N^*(1535)$ are also marked by
     filled diamonds.}
  \label{fig:switching}
\bigskip
\end{figure}

Now, we  compare the position of the Roper resonance ($N'$)
and the negative parity nucleon ($N^*$).
Plotted in Fig.~\ref{fig:switching} are
the masses of these resonances as well as the nucleon mass  
as a function of the pion mass squared in the physical unit.
All the points  are evaluated by the MEM analysis.  The
chiral extrapolations with the curve fit are also shown
together with the experimental values.
We find that the $N'$ and $N^*$ states
are almost degenerate within statistical errors 
in a  wide range of quark masses, $0.61 < M_{\pi} < 1.22 $ GeV,
although the central values of the $N'$ is slightly higher than the $N^*$
in this interval.
In the chiral limit, we have the following mass ratios:
%
%
\begin{eqnarray}
  \hspace{20mm}M_{N'}/M_N  &=& 1.64(35) 
  \hspace{4mm}(\mathrm{Expt.}\sim 1.53),  \nonumber \\
  \hspace{20mm}M_{N^*}/M_N &=& 1.48(22)
  \hspace{4mm}(\mathrm{Expt.}\sim 1.63). 
  \label{eqn:expt}
\end{eqnarray}
Our result here is in contrast to what is expected in the 
phenomenological quark models as mentioned in the Introduction,
but is consistent with
the experimental value. To make a firm conclusion on the 
precise level ordering between the $N'$ and $N^*$ states, we
need to increase the number of gauge configurations
to reduce the statistical errors and also need to
make simulations with smaller quark masses.
 
In relation to the physics
in the vicinity of the chiral limit,
we comment on the quenched artifact due to the $\eta'N$ ghost state
in the nucleon correlator. 
It is pointed out in \cite{Mathur:2003zf} that this ghost state
affects the nucleon correlator significantly for $M_\pi  \simle 0.3$ GeV.
If such effect is significant, the spectral function becomes
negative and the standard MEM analysis
assuming the positivity of the SPF is invalidated.  
In the present study, however, simulations
are  performed in the region $M_\pi=0.6-1.2$ GeV, 
and we have checked explicitly that the $\eta'N$ threshold is higher 
than the masses of the $N$, $N'$ and $N^*$ states in the this
interval.
 

\section{Bound state of doublers}
\label{sec:Bound_state_of_doublers}

  In this section, we discuss 
  the high-lying peaks of the SPF which
  could be the bound states containing Wilson doublers.
  This possibility was first pointed out by 
  CP-PACS collaboration for the meson spectra~\cite{Yamazaki:2001er}. 
  Since the mass of the Wilson doublers
  is inversely proportional to the lattice spacing $a$, 
  studying the peak positions of SPF as a function of $\omega a$
  provides us with a useful information on 
  the doubler bound states.
 
 %
 %
\begin{figure}[tbhp]
\bigskip
  \begin{center}
  \includegraphics[height=65mm]{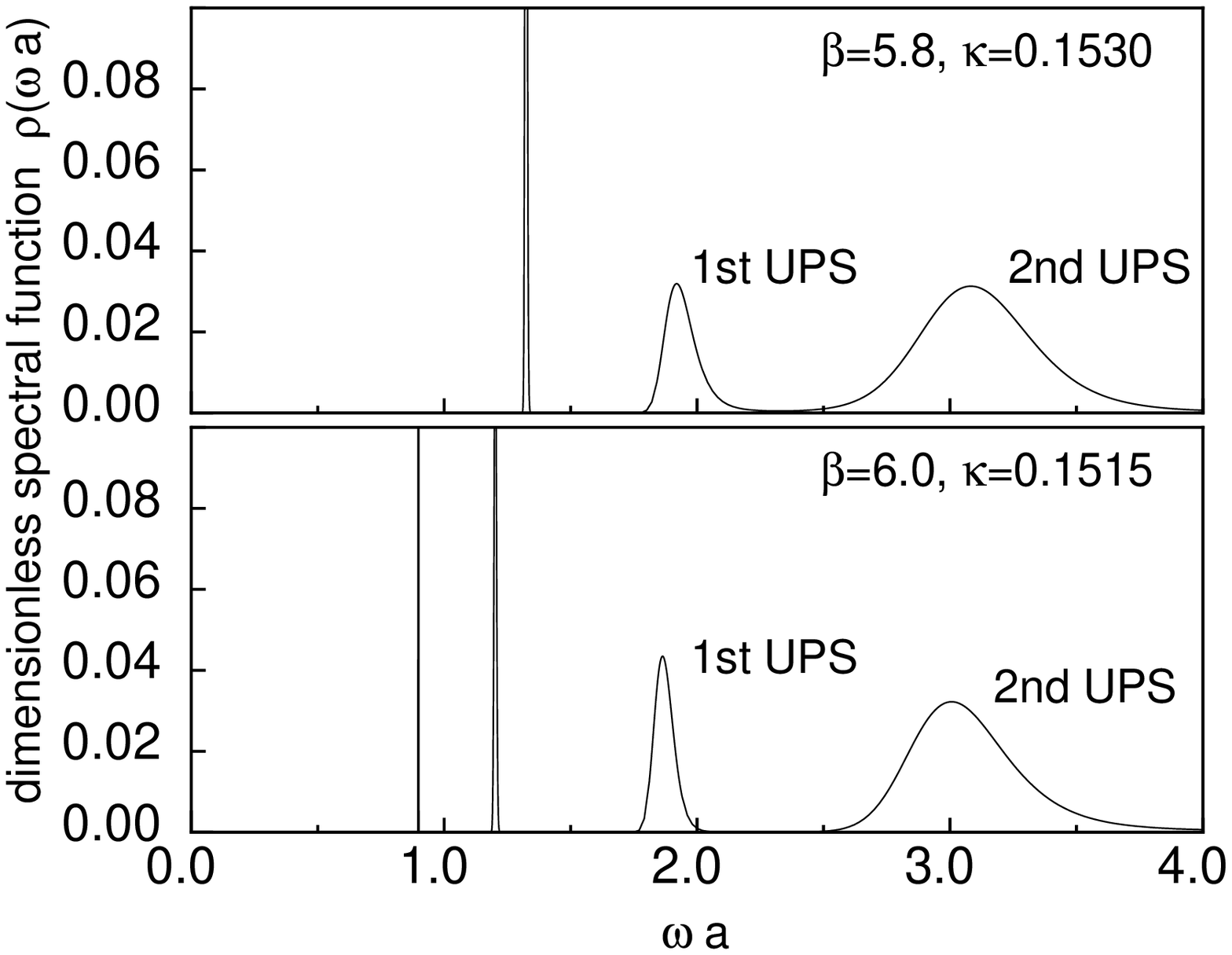}
  \hspace{0mm}
  \includegraphics[height=65mm]{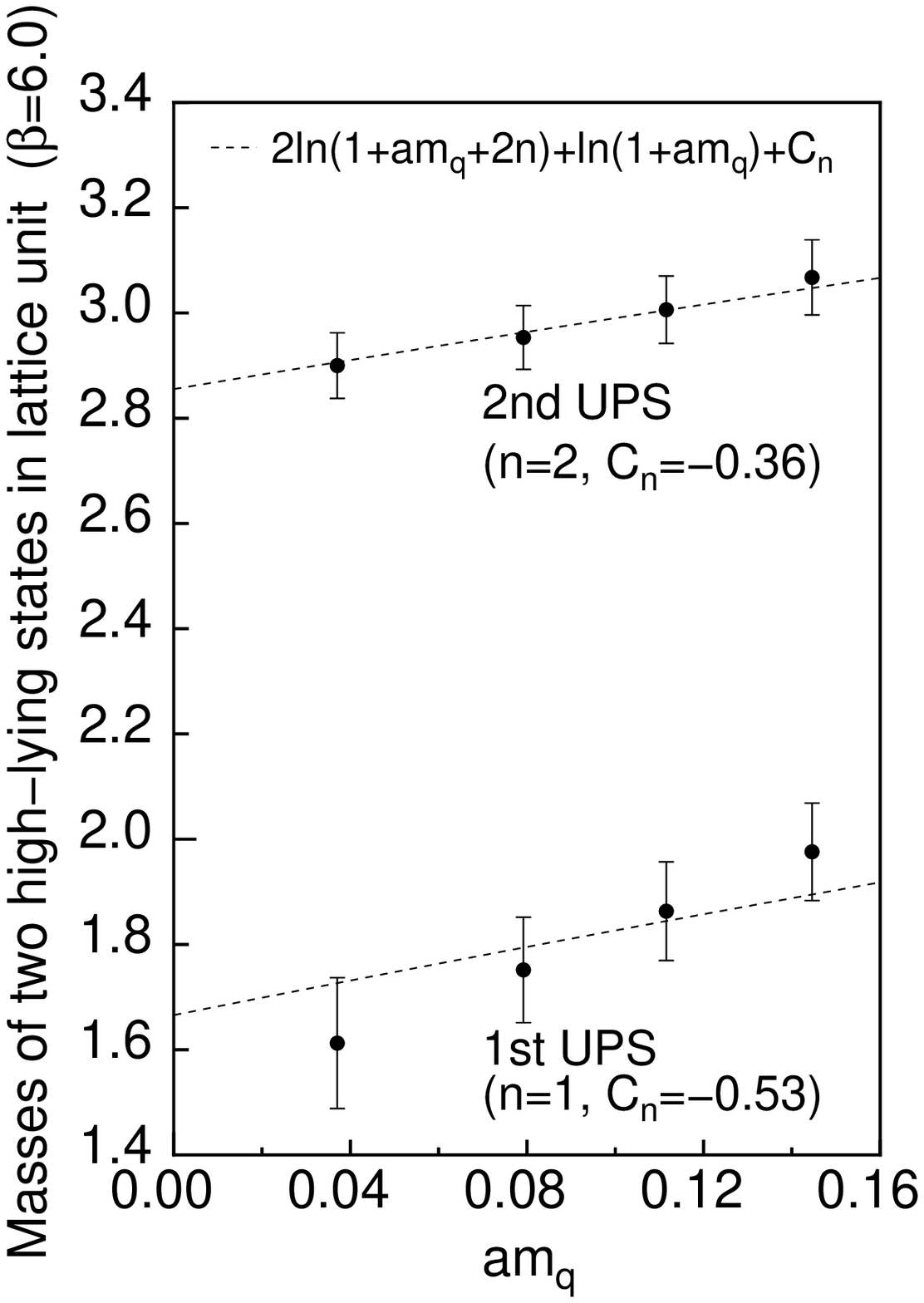}
  \end{center}
  \caption{Left panel: A comparison of the SPF in the positive parity channel 
  at two different couplings, $\beta=5.8,~6.0$. 
  The hopping parameter is taken as $\kappa=0.1530$ ($\kappa=0.1515$)
  for $\beta=5.8$ ($\beta=6.0$) so that we have $M_{\pi}\simeq 1$ GeV
  in both cases. Right panel: 
  The positions of the two high-lying peaks, labeled by 1st UPS (first unphysical state)
  and 2nd UPS,
  are plotted against the bare quark-mass $am_q=(1/\kappa-1/\kappa_c)/2$. 
  Results of fitting the data using a mass formula given in the text
  are shown by the dotted lines. }
  \label{fig:DOUBLER_COMPARISON}
\bigskip
\end{figure}

 Shown in the left panel of Fig.~\ref{fig:DOUBLER_COMPARISON}
 are the SPFs in the positive parity nucleon channel  
 for $M_\pi\simeq 1$ GeV with $\beta=5.8$ (upper figure)
 and $\beta=6.0$ (lower figure). The lattice volume is
 taken to be $La \simeq 3.0$ fm.
 The MEM analysis is performed with $t_\mathrm{max}=14$ 
 for $\beta=5.8$ and 20 for $\beta=6.0$. 
 These values of $t_\mathrm{max}$ are chosen so that
 the resulting images of SPF are stable enough. 
 The number of configurations in both cases  is $N_\mathrm{conf}=200$. 
 
 Consider the two high-lying peaks whose positions in terms of
 $\omega a$ are almost independent of the lattice spacing $a$.
 Insensitivity of the peak positions in terms of $\omega a$ is
 already an indication that they contain Wilson doublers,
 the pole mass of a free Wilson doubler is given by 
 $am^{\rm pole}_{n\neq 0}=\mathrm{ln}(1+am_{q}+2n)$  
 with $n$ being an integer. The bare quark-mass $m_q$ is defined 
 by $am_{q}\equiv(1/\kappa-1/\kappa_c)/2$.
 Note that baryons at rest composed of Wilson doublers
 need to contain two doublers to satisfy the 
 momentum conservation.  
Then one may write a simple
 mass formula for two high-lying peaks as 
%
%
\begin{eqnarray} 
 aM_{ n\neq 0}= 2\ \mathrm{ln}(1+am_q+2n)+\mathrm{ln}(1+am_q)+C_{n},
 \label{eq:doubler}
\end{eqnarray}
where $C_{n}$ corresponds to a binding energy
between the two doublers and a physical quark. 
As far as  $n \neq 0$,  $aM_n$ depends only weakly on the  bare quark-mass $m_q$.

In the right panel of Fig.~\ref{fig:DOUBLER_COMPARISON},
the positions of the highest two peaks for $\beta=6.0$
are plotted as a function of the bare quark-mass, $am_q$.
Results of fitting the data by Eq.(\ref{eq:doubler})
are shown by the dotted lines. 
We find that the mass formula reproduces the positions and the 
$m_q$ dependence of the highest two peaks well.
As a result of this quantitative analysis, it is quite
likely that the highest two peaks are the bound states
containing Wilson doublers which eventually decouple
from the physical spectrum in the continuum limit, $a \rightarrow 0$.

 It is worth mentioning that any prominent structure of SPF
 in the higher frequency region ($\omega \gg 1/a$)
 is not found in the 4-dimensional $O(4)$ $\sigma$ model 
\cite{Yamazaki:2002ir} and
 the 3-dimensional four-fermion model 
with staggered fermions \cite{Allton:2002mv}. These 
 observations support that our observed high-lying peaks
 far beyond cutoff scale $1/a$ are associated with Wilson doublers.

\section{Comparison to other approaches}
\label{sec:Comparison_with_other_results}

%
%
\begin{table}[tbhp]
\bigskip
\caption{Summary of the simulation parameters and methods
  in  previous analyses together with the current analysis. 
  The fermion action, the lattice spacing, the pion mass, 
  the spatial size, 
   the boundary condition in time
  and the method of analysis are tabulated.
  CID stands for the chirally improved Dirac operator, CCF for 
  the constrained curve fitting. 
  Note that the physical scales are set 
  by the pion-decay constant $f_\pi$ in Ref. \cite{Mathur:2003zf}, 
  and by the Sommer parameter $r_0$ in other references. }
%
%
%

%
%
\begin{center}
\begin{tabular}{ccccc|lr}
\hline
\hline
fermion & $a^{-1}$ (GeV) & $M_\pi$ (GeV) & $La$ (fm) & B.C. & 
method (Ref.)\\
\hline
DWF    & 2.1 & $0.57-1.43$ & 1.5      & P+AP & $2\times2$ correlator~\cite{Sasaki:2001nf} \\
clover & 2.9 & $0.51-1.08$ & 1.6      & P & modified correlator~\cite{Guadagnoli:2004wm} \\
CID & 1.3 & $0.45-0.87$ & 1.8      & AP   & $4\times4$ correlator~\cite{Burch:2004he}\\
overlap& 1.0 & $0.18-0.87$ & 2.4, 3.2 & AP  & CCF~\cite{Mathur:2003zf}\\ 
Wilson & 2.1 & $0.61-1.22$ & 1.5, 3.0 & P+AP  & MEM (this work)\\
\hline
\hline
\end{tabular}
\end{center}
\label{table:comparison_other_group}
\bigskip
\end{table}

 In this section, we  briefly make a comparison
 among  different approaches applied to analyze the excited nucleons.  (See
 Table \ref{table:comparison_other_group}.) 
 The methods to extract the
 excited spectra may be  classified into three categories:
 (i) Diagonalization method using correlation 
 matrix~\cite{Sasaki:2001nf,Burch:2004he}, 
 (ii) Subtraction method using modified correlator~\cite{Guadagnoli:2004wm} 
 and (iii) Bayesian methods such as  
 the maximum entropy method (\cite{Sasaki_Roper} and the present work)
 and the constrained
 curve fitting (CCF) \cite{Mathur:2003zf}.
 
 In (i), one starts with the correlation matrix, 
 $C_{ij}(t)=\langle O_i(t)\overline{O_j}(0) \rangle$, 
 and solve the generalized eigenvalue problem $C(t){\vec v}=\lambda(t,t_0)C(t_0){\vec v}$
 where $t_0$ is fixed and $t>t_0$. The $k$-th largest eigenvalue of the transfer matrix 
 $\lambda(t,t_0)$ may be given by $\lambda^{(k)}(t,t_0)=e^{-(t-t_0)M_{k}}$ 
 with appropriate choices of $t$ and $t_0$~\cite{Luscher:1990ck}. 
 In Ref. \cite{Sasaki:2001nf}, 
 an unconventional operator, 
 $\varepsilon_{abc}[u_a^{\rm T}(x)Cd_b(x)]\gamma_5u_c(x)$  
 is utilized to construct the $2\times 2$ correlation matrix 
 in addition to the conventional one as Eq.(\ref{eq:conv-N}).  
 On the other hand, in Ref. \cite{Burch:2004he}, they build up the correlation
 matrix with four operators, which correspond to 
 four possible combinations to construct the nucleon operator
 from two sets of the smeared quark propagator.
 In (ii),
 a modified correlator, $\hat{G}(t)\equiv G(t+1)G(t-1)-G^2(t)$,
 is investigated instead of the original 
 correlator $G(t)$, so that 
  the sum of the first two states, $M_0+M_1$, 
  governs the leading-exponential decay of the modified correlator
 as $\hat{G}(t)=2\sum_{n>0}A_0A_n
 [\cosh(M_n-M_0)-1]~\mbox{e}^{-(M_0+M_n)t}$  \cite{Guadagnoli:2004wm}. 
 In (iii), prior knowledge of the correlator is considered in addition to the 
 Monte Carlo data.
 In MEM \cite{MEM_lQCD}, the prior knowledge is introduced 
 in the entropic form, while in CCF, it is
 introduced in the gaussian form \cite{Lepage:2001ym}.

 In Refs.\cite{Sasaki:2001nf,Burch:2004he,Guadagnoli:2004wm}
 (the first three works in Table \ref{table:comparison_other_group}),
 degeneracy between the $N'$ and $N^*$ states has
 not been observed within the range of the quark masses
 where simulations are performed.
 On the other hand, in  Ref.\cite{Mathur:2003zf} and
 the present work (the last two works 
 in Table~\ref{table:comparison_other_group}),
 an approximate degeneracy of the $N'$ and
 the $N^*$ are seen  within the
 statistical errors, although it is not yet clear where and how the
 level switching between the $N'$ and $N^*$ states really takes
 place as a function of the quark mass  in our study.
 One of the major differences between the first three works and
 the last two works is the spatial lattice size.

  As we have discussed in Section \ref{sec:NSTR_and_Roper}
  and will be reported in detail in a separate publication \cite{in_preparation},
  our simulation with a smaller lattice ($La \simeq 1.5 $fm)
  show that there is a sizable upward shift
  especially in the $N'$ channel for $M_{\pi}\simeq 0.6$ and $0.9$ GeV.
  This lattice artifact increases the $N'-N^*$ difference
  even for rather heavy quarks
  and may partly explain the
  reason why degeneracy have not seen in first three works.
    
\section{Conclusion}
\label{sec:Conclusion}

In this article, we have studied the mass spectra of excited nucleons 
in the quenched lattice QCD simulations. 
We have focused our attention on the level ordering 
between $N'(1440)$ (the nucleon excited state in the positive-parity channel) 
and $N^*(1535)$ (the nucleon excited state in the negative-parity channel) 
which has been a long standing puzzle. 
 Indeed, it is difficult to be explained 
by phenomenological quark models in which
the Roper resonance $N'$ has much higher mass than the $N^*$. 

To attach the problem with a firm footing,
we have utilized a technique to combine quark propagators 
with periodic and anti-periodic boundary conditions in time to 
eliminate
the contamination of the opposite 
parity states in the baryonic correlator  up to the first wrap-round effect.
Also, we have used the maximum entropy method (MEM) 
to reconstruct the spectral function
which contains not only the information of the ground state 
but also that of the excited states. 
Since the finite size effect is shown to be
severe for excited nucleons \cite{Sasaki:2005ug},
we have taken a large lattice size
($La\simeq 3$ fm with $\beta=6.0$).

 With all the above, we have extracted the masses
 of  the $N$, $N'$ and $N^*$ states for
 $M_{\pi}=0.61-1.22$ GeV, which are summarized in 
 Fig.\ref{fig:switching}.
 Our results show that 
 the masses of  the $N'$ and $N^*$ states are
 approximately degenerate within statistical errors for
 wide range of quark masses, although the 
 central value of the $N'$ is slightly higher than the $N^*$.
 This tendency is valid also at the physical point
 after making chiral extrapolation with a curve fit, Eq.(\ref{eqn:expt}). 
 The latter result is in contrast to 
 the predictions of the phenomenological quark models 
 but is consistent with similar finding in Ref.~\cite{Mathur:2003zf} 
 which employs
 different lattice action and different way of data
 analysis.
 Whether level switching between the $N'$ and $N^*$ states takes place
 at  certain quark mass is an open question left for future studies.
 We have also studied
 the high-lying peaks of the SPF by carrying
 out simulations on a coarse lattice ($\beta=5.8$)
 with approximately the same lattice volume.  From the analysis
 of the $a$-dependence and the quark-mass dependence
 of these peaks, we found 
 a firm evidence that the high-lying peaks correspond
 to the bound states composed of two Wilson doublers and
 one normal quark.  

 There are various directions to be explored in future studies:
 To increase the statistics to pin down the precise
 location of the level switching; To generalize the MEM 
 to treat the negative SPF originating from the 
 ghost $\eta'N$ state; To make systematic studies of
 the other excited baryons and mesons. 
 The nucleon excited states at finite temperature are also interesting to be studied
 in connection with realization of chiral symmetry in baryons~\cite{DeTar:1988kn}.
 The MEM analysis also may be helpful to disentangle the pentaquark signal from 
 a tower of the $KN$ scattering states~\cite{Sasaki:2003gi}.

\section*{Acknowledgments}
It is a pleasure to acknowledge A. Nakamura and C. Nonaka 
for helping us develop codes for our lattice QCD simulations 
from their open source codes (Lattice Tool Kit~\cite{Choe:2002pu}).
This work is supported by the Supercomputer Projects No.102 (FY2003) 
and No.110 (FY2004) of High Energy Accelerator Research Organization (KEK). 
T.H. and S.S. thank for the support by Grants-in-Aid 
of the Japanese Ministry of Education, Culture, Sports,
Science and Technology (No.15540254 and No.15740137).


\end{document}